\documentclass[prb,twocolumn,longbibliography,superscriptaddress,preprintnumbers]{revtex4-2}
\usepackage{bm}
\usepackage{dsfont}
\usepackage{graphicx}
\usepackage{epsfig}
\usepackage{epstopdf}
\usepackage{balance}
\usepackage[dvipsnames]{xcolor}
\usepackage{calc}
\usepackage{natbib}

\usepackage{newunicodechar} 
\usepackage{lipsum}
\usepackage[version=3]{mhchem} 
\usepackage{color}
\usepackage{soul}
\usepackage[etex=true,export]{adjustbox}
\usepackage{makecell}
\usepackage{multirow}
\usepackage{tabularx,multirow,array,diagbox}
\usepackage{adjustbox}
\usepackage{booktabs}
\makeatletter
\renewcommand{\maketag@@@}[1]
{\hbox{\m@th\normalsize\normalfont#1}}%

\newcommand\bl[1]{\boldsymbol{#1}}
\makeatother

\usepackage[colorlinks,
            linkcolor=blue,
            anchorcolor=blue,
            citecolor=blue,
            urlcolor=blue]{hyperref}
  
\begin{document}

\title{Quantum Phases in Twisted Homobilayer Transition Metal Dichalcogenides}
\author{Bohao Li}
\affiliation{School of Physics and Technology, Wuhan University, Wuhan 430072, China}
\author{Wen-Xuan Qiu}
\affiliation{School of Physics and Technology, Wuhan University, Wuhan 430072, China}
\author{Fengcheng Wu}
\email{wufcheng@whu.edu.cn}
\affiliation{School of Physics and Technology, Wuhan University, Wuhan 430072, China}
\affiliation{Wuhan Institute of Quantum Technology, Wuhan 430206, China}
\author{A. H. MacDonald}
\email{macd@physics.utexas.edu}
\affiliation{Department of Physics, University of Texas at Austin, Austin, Texas 78712, USA }

\begin{abstract}
    Twisted homobilayer transition metal dichalcogenides—specifically twisted bilayer MoTe$_2$ and twisted bilayer WSe$_2$—have recently emerged as a versatile platform for strongly correlated and topological phases of matter. These two-dimensional systems host tunable flat Chern bands in which Coulomb interactions can dominate over kinetic energy, giving rise to a variety of interaction-driven phenomena. A series of groundbreaking experiments have revealed a rich landscape of quantum phases, including integer and fractional quantum anomalous Hall states, quantum spin Hall states, anomalous Hall metals, zero-field composite Fermi liquids, and unconventional superconductors, along with more conventional topologically trivial correlated states including antiferromagnets. This review surveys recent experimental discoveries and theoretical progress in understanding these phases, with a focus on the key underlying mechanisms—band topology, electron interactions, symmetry breaking, and charge fractionalization. We emphasize the unique physics of twisted TMD homobilayers in comparison to other related systems, discuss open questions, and outline promising directions for future research.
\end{abstract}

\maketitle

\section{Introduction}

The field of moir\'e materials \cite{Bistritzer2011moire}
has sparked a revolution in condensed matter physics by 
providing a new generation of quantum materials—one in which properties can be 
flexibly tuned by using different two-dimensional crystals with different 
lattice constant mismatches, by changing interlayer twists, and most 
crucially by changing gate voltages {\em in situ} to induce continous property evolution. 
In semiconductor and semimetal hosts, moir\'e superlattices yield emergent periodic 
Hamiltonians—artificial large-lattice-constant crystals—with 
Bloch states and, frequently, flat electronic bands. The large periodicities allow gates to change the electron filling per moir\'e unit cell by $\sim 10$, effectively moving across the columns of a chemical periodic table.  Each integer filling corresponds to a distinct entry in a {\em moir\'e periodic table} and intermediate filling factors correspond to disorder-free doping.  

The emergence of moir\'e materials as 
an important topic in condensed matter physics was catalyzed by the discovery of correlated insulators and superconductivity in magic-angle twisted bilayer graphene (TBG) \cite{Cao2018Correlated,Cao2018Unconventional}, but the field has since 
expanded to include a much broader range of moir\'e materials \cite{Andrei2020Graphene,Balents2020Superconductivity,Mak2022semiconductor,Nuckolls2024microscopic}, including some based on transition metal dichalcogenides (TMDs). Among these, 
twisted TMD homobilayers—specifically twisted bilayer MoTe$_2$ ($t$MoTe$_2$) and twisted 
bilayer WSe$_2$ ($t$WSe$_2$) in the $R$-stacking configuration—have emerged as a 
particularly fertile ground for the discovery of new strongly correlated and topologically 
nontrivial phases of matter, including unusual examples of superconductivity. 
Monolayers of group-VIB TMDs, with chemical composition $MX_2$ ($M=$Mo, W and $X$=S, Se, Te), are semiconductors in the $H$ phase. In these monolayers, valence band maxima are located at the two inequivalent corners of the hexagonal Brillouin zone, known as $\pm K$ valleys. Due to the absence of inversion symmetry, spin-orbit coupling generates a pronounced valley-contrasting Ising spin splitting in the valence bands, leading to an effective spin-valley locking \cite{Xiao2012Coupled}. When hole doping places the chemical potential near the valence band maxima, the low-energy carriers can be modeled as massive particles possessing a valley pseudospin degree of freedom, which effectively behaves as the physical spin. This behavior is distinct from that of monolayer graphene, in which
the low-energy carriers are described by massless Dirac fermions with both spin and valley degrees of freedom. 

These intrinsic differences at the monolayer level result in qualitatively distinct physical properties in the twisted bilayer counterparts. In TMDs, spin-valley locking depletes
the low-energy degrees of freedom, giving rise to moir\'e bands with a simplified flavor structure. The massive nature of low-energy carriers leads to moir\'e bands with narrow bandwidths over a broad range of twist angles, in sharp contrast to the magic-angle condition required to achieve flat bands in TBG. Furthermore, the lack of twofold rotational symmetry about the out-of-plane $\hat{z}$ axis in twisted homobilayer TMDs permits valley-contrasting Berry curvatures and enables the realization of topological moir\'e bands driven by interlayer hybridization. These characteristics are elegantly
captured by a simple continuum moir\'e Hamiltonian \cite{Wu2019Topological}, which reveals that particles move within a background layer pseudospin skyrmion texture, providing a real-space view of the emergence of moir\'e band topology. The top two moir\'e valence bands can carry opposite Chern numbers, realizing an effective Haldane model \cite{Haldane1988Model} in a single valley and a Kane-Mele model \cite{Kane2005Quantum} when both valleys are considered.

Due to the narrow moir\'e band widths, electron interaction effects are significantly enhanced. Early theoretical studies proposed that at hole filling factor $\nu_h=1$ (one hole per moir\'e unit cell), the system could realize an integer quantum anomalous Hall state driven by spontaneous valley polarization \cite{Wu2019Topological}. Going beyond integer fillings, numerical studies based on exact diagonalization have further predicted fractional quantum anomalous Hall insulators at fractional hole fillings in $t$MoTe$_2$ \cite{Li2021Spontaneous}.  The possibility of realizing such fractionalized states in twisted homobilayer TMDs was further supported by the theoretical observation that the moir\'e  flat bands in this system can have relatively uniform Berry curvature distribution—a key condition for stabilizing fractional Chern insulators \cite{Devakul2021}.

The potential of this platform was first revealed by a pioneering transport study of $t$WSe$_2$, which  uncovered a topologically trivial correlated insulating state at $\nu_h=1$ indicating the strongly interacting nature of the system \cite{Wang2020Correlated}. A major turning point in this field came with the discovery of both integer and fractional quantum anomalous Hall insulators in $t$MoTe$_2$ \cite{Cai2023,Zeng2023,Park2023,Xu2023Observation}. Although integer quantum anomalous Hall insulators had already been observed in various material platforms \cite{Chang2013Experimental,Zhao2020Tuning,Deng2020Quantum,Liu2020Robust,Zhang2019Topological,Li2019Intrinsic,Otrokov2019Prediction,Sharpe2019Emergent,Serlin2020Intrinsic,Chen2020Tunable,Polshyn2020Electrical,Polshyn2022Topological,Li2021}, $t$MoTe$_2$ was the first system to exhibit fractional quantum anomalous Hall effects at zero external magnetic field \cite{Park2023,Xu2023Observation}, marking a significant milestone in the pursuit of fractionalized topological phases.

\begin{figure*}[t]
    \includegraphics[width=2.\columnwidth]{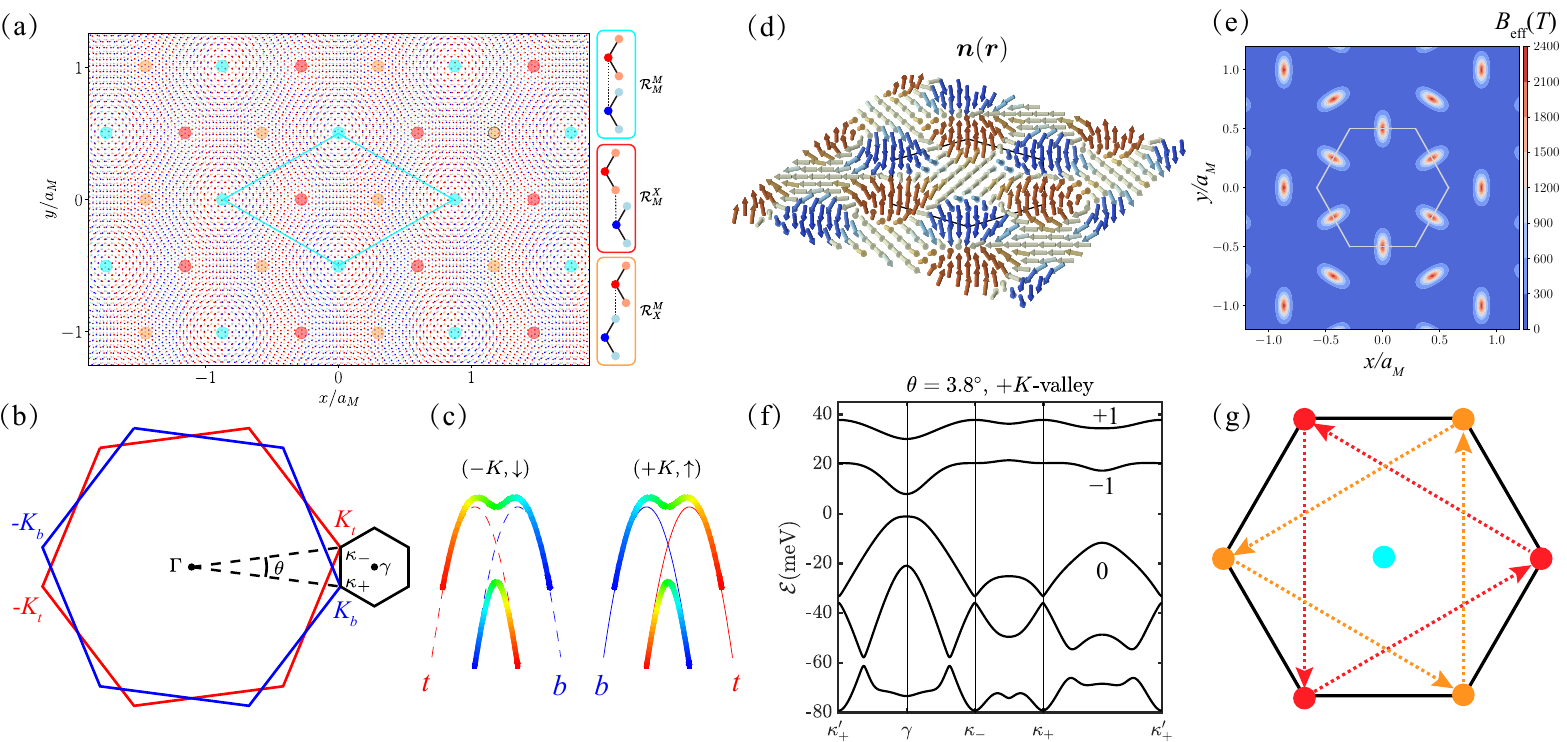}
    \caption{Single-particle physics in twisted homobilayer TMDs. (a) Moir\'e superlattices of $R$-stacked twisted homobilayers. The dots identify the high symmetry positions $\mathcal{R}_M^M$, $\mathcal{R}_X^M$, and $\mathcal{R}_M^X$, and the solid lines outline a single moir\'e unit cell. (b) Brillouin zones of the bottom (blue) and top (red) layers in a twisted bilayer, and the moir\'e Brillouin zone (black). (c) Moir\'e band states near the valence band maxima of two valleys.  
    The color shading indicates the degree of 
    layer polarization of the corresponding layer spinors (bottom=blue and top=red).
    (d) Map of the skyrmion field $n(\bl r)$ and (e) the corresponding effective magnetic field $B_{\mathrm{eff}}(\bl r)$ of $t$MoTe$_2$ at $\theta= 3.8^\circ$. The black hexagon in (d) (the white hexagon in (e)) is the Wigner-Seitz cell of the moir\'e superlattice. (f) Moir\'e band structure of $t$MoTe$_2$ at $\theta= 3.8^\circ$ calculated from the continuum model. The numbers $(+1,-1,0)$ are the Chern numbers of the first three moir\'e valence bands. (g)  The honeycomb lattice formed by $\mathcal{R}_X^M$, and $\mathcal{R}_M^X$ sites.  Red and orange dashed arrows indicate the phase pattern of next-nearest-neighbor hopping in the effective Haldane model for the first two bands in (f). Panels (a)-(c) are adapted from Ref.~\cite{Pan2020Band}.}  
    \label{fig:1}
\end{figure*}

Beyond these advances, twisted homobilayer TMDs have 
been shown to host a rich variety of quantum phases in a phase diagram tuned by 
carrier-density, displacement-field, and twist angle \cite{Wang2020Correlated,Park2023,Anderson2024Trion,kang2024evidence,kang2024evidence,Xu2025Interplay,Xu2025signatures,Guo2025Superconductivity,Xia2025Superconductivity,Guo2025Superconductivity,Xia2025Superconductivity}.  States discovered to date 
include integer and fractional quantum spin Hall insulating states \cite{kang2024evidence,kang2024evidence,Xu2025Interplay}, anomalous Hall metals \cite{Park2023}, zero-field composite Fermi liquids \cite{Park2023,Anderson2024Trion}, unconventional superconductors  \cite{Xu2025signatures,Guo2025Superconductivity,Xia2025Superconductivity}, and topologically trivial correlated phases such as intervalley antiferromagnets \cite{Wang2020Correlated,Guo2025Superconductivity,Xia2025Superconductivity}. 
The emergence of these diverse quantum states has attracted broad interest and opened new research directions within the topological physics, superconductivity, and quantum materials fields.  

In this paper, we present a comprehensive review of the emergent quantum phases in twisted TMD homobilayers. We start by examining the moir\'e band structures, emphsazing their topological properties which lay the foundation for a deeper investigation of the resulting quantum phases. We then summarize key experimental findings, highlighting the experimental signatures of various quantum phases when examined with different probes.  Following this, we explore in detail the integer and fractional topological states and the superconducting 
states, connecting them to theoretical frameworks that emphasize correlation effects within flat Chern bands. Throughout the review, we highlight the unique characteristics of twisted homobilayer TMDs in comparison to other related systems, and conclude with a discussion of open challenges and promising directions for future research.

\section{Topological Moir\'e Bands}

In twisted TMD homobilayers (both layers from the same material), configurations with twist angles $\theta$ near $0^\circ$ and $180^\circ$ are physically distinct. This is because the individual TMD monolayers lack $C_{2z}$ symmetry (i.e., twofold rotation 
symmetry around the out-of-plane $\hat{z}$ axis). 
These two cases correspond to different stacking types, referred to as 
$R$  and $H$ respectively. The labels are inherited from terminology used to describe the corresponding untwisted bilayers; $R$ (rhombohedral) refers to the aligned configuration and $H$ (hexagonal) to the antialigned configuration. In the $H$ stacking of twisted bilayers, opposite layers have opposite spins and time-reversed atomic-orbital wavefunctions in a given valley.
The spin mismatch suppresses interlayer tunneling, effectively decoupling the two layers at low energies at the single-particle level. Consequently, $H$ stacking can realize a bilayer generlized Hubbard model in which layer and valley degrees of freedom together give rise to an approximate SU(4) symmetry in a valley/layer flavor space \cite{Wu2019Topological,Zhang2021SU(4),Xu2022Atunable}. In contrast, in $R$ stacking momentum-aligned valleys in the two layers carry the same spin and have the same atomic orbital, allowing for stronger interlayer tunneling that imposes qualitatively different physics. 

It is important to note that the location of the valence band maximum in twisted TMD homobilayers can vary depending on the material. For examples, the valence band maxima shift to the Brillouin zone center (i.e., the $\Gamma$ valley \cite{Angeli2021}) in twisted bilayers of $t$MoS$_2$, $t$WS$_2$, and $t$ MoSe$_2$, but remain in the $\pm K$ valleys 
in $t$WSe$_2$ and $t$MoTe$_2$. These two cases have qualitatively distinct electronic 
structure and low-energy physics. 
In this review article, we focus on $t$MoTe$_2$ and $t$WSe$_2$ in the $R$-stacking configuration, in which the low-energy physics is controlled by $\pm K$ valley band
states with strong interlayer hybridization.

The moir\'e superlattices formed in an $R$-stacked twisted homobilayer is illustrated in Fig.~\ref{fig:1}(a). For small $\theta$, the supelattices have a moir\'e period $a_M \approx a_0/\theta$, where $a_0$ is the in-plane constant of the monolayer and $\theta$ is the 
twist angle in radians.  The twisted bilayer possesses $D_3$ point-group symmetry generated by a threefold rotation $C_{3z}$ around the  $\hat{z}$ axis and a twofold rotation $C_{2y}$ around the in-plane $\hat{y}$ axis that exchanges the two layers. Within a moir\'e unit cell, there are three positions invariant under $C_{3z}$: $\mathcal{R}_{M}^M$, $\mathcal{R}_{M}^X$, and $\mathcal{R}_{X}^M$. Here  $\mathcal{R}_{\alpha}^{\beta}$ denotes the local atomic registry where  the $\alpha$ atom in the bottom TMD layer is vertically aligned with the $\beta$ atom in the top layer. Under the $C_{2y}$ operation, the $\mathcal{R}_{M}^M$ sites remain invariant, while $\mathcal{R}_{M}^X$ and $\mathcal{R}_{X}^M$ sites are exchanged.
  
In momentum space, states in $+K$ and $-K$ valleys can be treated separately in the single-particle Hamiltonian, since they are separated by a large momentum when $\theta$ is small [Fig.~\ref{fig:1}(b)]. The two valleys are related by the $C_{2y}$ symmetry and by time-reversal symmetry $\mathcal{T}$. For definiteness, we describe the single-particle physics in the $+K$ valley, with the understanding that the $-K$ valley states can be obtained via the action of $C_{2y}$ or $\mathcal{T}$ symmetries. The low-energy valence states in the $+K$ valley carry spin up ($\uparrow$) due to spin-valley locking and are described by the following moir\'e Hamiltonian\cite{Wu2019Topological},  
	\begin{equation}
	\mathcal{H}_\uparrow=\begin{pmatrix}
	-\frac{\hbar^2 (\hat{\bm{k}}-\bm{\kappa}_+)^2}{2m^*}+\Delta_{+}(\bm{r}) & \Delta_{\text{T}}(\bm{r})\\
	\Delta_{\text{T}}^\dagger(\bm{r}) & -\frac{\hbar^2(\hat{\bm{k}}-\bm{\kappa}_-)^2}{2m^*}+\Delta_{-}(\bm{r})
	\end{pmatrix},
	\label{eq:moire}
	\end{equation}
where the $2\times 2$ matrix is in the layer pseudospin space, the diagonal terms are intralayer Hamiltonians, and the off-diagonal terms describe the interlayer tunneling. Here, $\bm r$ and $\hbar \hat{\bm k}$ are, respectively the position and momentum operators, and $m^*$ is the valence band effective mass. The layer-dependent momentum offsets $\bm{\kappa}_{\pm}=[4\pi/(3a_M)](-\sqrt{3}/2,\mp 1/2)$ capture the relative momentum shift of valence band maxima  between the two twisted layers [Fig.~\ref{fig:1}(c)]. The layer-dependent moir\'e potential $\Delta_{\pm}(\bm{r})$ is given by
	\begin{equation}
	\Delta_{\pm}(\bm{r}) = 2 V \sum_{j=1,3,5}^{}\cos(\bm{g}_j\cdot \bm{r} \pm \psi),
    \label{Deltapm}
	\end{equation}
where $V$ and $\psi$ respectively characterize the amplitude and spatial pattern of the moir\'e potential . The interlayer tunneling $\Delta_{\text{T}}(\bm{r})$ is fully
characterized by a strength parameter $w$:
	\begin{equation}
	\Delta_{\text{T}}(\bm{r}) = w (1+e^{-i \bm{g}_2 \cdot \bm{r}}+e^{-i \bm{g}_3 \cdot \bm{r}}).
    \label{DeltaT}
	\end{equation}
In Eqs.~\eqref{Deltapm} and \eqref{DeltaT}, $\boldsymbol{g}_{j}=\frac{4\pi}{\sqrt{3}a_M}[\cos\frac{\pi(j-1)}{3},\sin\frac{\pi(j-1)}{3}]$ are moir\'e reciprocal lattice vectors in the first momentum shell.  

The continuum Hamiltonian $\mathcal{H}_\uparrow$ acts on envelope
functions that vary on the moir\'e length scale and assumes that interlayer 
tunneling is accurately local on that scale. 
Moir\'e translation symmetry with period $a_M$ justifies the reciprocal lattice
Fourier expansions of $\Delta_{\pm}$ and $\Delta_{T}$, 
and the large ratio of inter-layer to intra-layer atomic distances justifies the blue truncation
at the leading harmonics.  The number of independent model parameters is 
further limited by applying $C_{3z}$ and $C_{2y} \mathcal{T}$ symmetries.
Because of the combined $C_{2y} \mathcal{T}$ symmetry, $\Delta_{+}(x,y)=\Delta_{-}(-x,y)$. The interlayer tunneling term $\Delta_{\text{T}}(\bm{r})$ has zeroes at the $\mathcal{R}_{M}^X$ and $\mathcal{R}_{X}^M$ positions that are protected by the $C_{3z}$ symmetry. An important feature of  $\Delta_{\text{T}}(\bm{r})$ is its complex spatial dependence on $\bm{r}$, which originates from the spatially-varying phase factors of the Bloch states at the Brillouin-zone corners. We note that  $\mathcal{H}_\uparrow$ in Eq.~\eqref{eq:moire} only includes quadratic kinetic energy terms and lowest-harmonic expansions in the potential terms. While higher-order terms are allowed and can be 
added when there is a motivation, $\mathcal{H}_\uparrow$ succintly captures the essential physics.

The symmetry properties of $\Delta_{\pm}(\bm r)$ and $\Delta_{\text{T}}(\bm r)$ imply a skyrmion lattice in real space, which is revealed by decomposing the moir\'e Hamiltonian using layer pseudospin Pauli matrices \cite{Wu2019Topological}.  A unitary transformation can be applied to $\mathcal{H}_\uparrow$ to make its form more symmetric, 
\begin{equation}
\begin{aligned}
  \mathcal{H}_{\uparrow}^{'} &=  U_{0}^\dagger(\boldsymbol r)\mathcal{H}_{\uparrow}U_{0}(\boldsymbol r),  \\
 &= -\frac{\hbar^2\hat{\boldsymbol k}^2}{2m^*}\sigma_0+\boldsymbol \Delta(\boldsymbol r)\cdot \boldsymbol\sigma+\Delta_0(\boldsymbol r)\,\sigma_0, 
\label{A1}
\end{aligned}
\end{equation}
where $U_0(\bm r)=\text{diag}(e^{i\boldsymbol \kappa_+\cdot \boldsymbol r},e^{i\boldsymbol \kappa_-\cdot \boldsymbol r})$, $\sigma_0$ is the identity matrix, and $\boldsymbol{\sigma}$ are Pauli matrices. The scalar potential $\Delta_0(\boldsymbol r)$ and the layer pseudospin  field $\boldsymbol\Delta(\boldsymbol r)$ are defined as
\begin{equation}
\label{Delta_de}
\begin{aligned}
\Delta_0(\boldsymbol r)= &\frac{\Delta_+(\boldsymbol r)+\Delta_-(\boldsymbol r)}{2},\\
\boldsymbol\Delta(\boldsymbol r)=  & [\text{Re}\,\widetilde\Delta_{\text{T}}^\dagger(\boldsymbol{r}),\text{Im}\,\widetilde\Delta_{\text{T}}^\dagger(\boldsymbol{r}),
\frac{\Delta_+(\boldsymbol r)-\Delta_-(\boldsymbol r)}{2}], \\
\widetilde\Delta_{\text{T}}^\dagger(\boldsymbol{r})=&e^{i(\boldsymbol \kappa_+-\boldsymbol   \kappa_-)\cdot \boldsymbol r}\Delta_{\text{T}}^\dagger(\boldsymbol r).
\end{aligned}
\end{equation}
The key observation is that the tunneling components of $\boldsymbol\Delta$ have vortex and antivortex textures around the $\mathcal{R}_{M}^{X}$ and $\mathcal{R}_{X}^{M}$ positions, while the $z$ component $\Delta_z$ takes opposite values at these two high-symmetry positions.  This spatial profile indicates that $\bm{\Delta}$ forms a skyrmion texture, characterized by the  winding number $N_w$,
	\begin{equation}\label{eq:windingnumber}
	\begin{aligned}
	N_w &\equiv\frac{1}{4\pi}\int_{\text{MUC}}d\bm{r} \; [\bm{n}\cdot(\partial_x \bm{n} \times \partial_y \bm{n})]\\
	&=\begin{cases}
	+1, V\sin\psi>0\\
	-1, V\sin\psi<0
	\end{cases},
	\end{aligned}
	\end{equation}
where the integral is over the moir\'e unit cell (MUC).
Here $\bm{n}$ denotes a unit vector in the direction of $\bm{\Delta}$. When $V\sin\psi > 0$, $\bm{n}$ points toward the north (south) pole at $\mathcal{R}_{M}^X$ ($\mathcal{R}_{X}^M$). This orientation reverses when $V\sin\psi < 0$, implying a change in sign of the 
the winding number $N_w$.

In the adiabatic limit where the particle's pseudospin  follows the skyrmion texture locally, there is an emergent orbital magnetic field $B_{\text{eff}}$ \cite{Wu2019Topological,Yu2019Giant,Pan2020Band,Nicolas2024Magic,li2025variational},  
	\begin{equation}\label{eq:bz}
	B_{\text{eff}}(\bm{r})=\frac{\hbar}{2e} \bm{n}\cdot(\partial_x \bm{n} \times \partial_y \bm{n}).
	\end{equation}
The effective magnetic flux produced by $B_{\text{eff}}$ over one MUC is quantized to $\pm \Phi_0$, where $\Phi_0=h/e$ is the magnetic flux quantum, following Eq.~\eqref{eq:windingnumber}. The spatial-averaged value of $B_{\text{eff}}$ has a magnitude of $\bar{B}=h/(e\mathcal{A}_0)$, where $\mathcal{A}_0=\sqrt{3}a_M^2/2$ is the area of one moir\'e unit cell (MUC). We can further define an effective magnetic length as  
\begin{equation}
    \ell_0=\sqrt{\frac{\hbar}{e\bar{B}}}=\sqrt{\frac{\sqrt{3}}{4\pi}}a_M\approx 0.37 a_M.
\end{equation}
For a typical moir\'e period $a_M$ of 5 nm, $\ell_0 \approx 1.85$ nm and  $\bar{B} \approx 191$ T—more than an order of magnitude larger than magnetic fields typically accessible in laboratory settings. Note that
$\bar{B}$ scales inversely with the square of the moir\'e period, i.e., $\bar{B} \propto 1/a_M^2 \propto \theta^2$.

The skyrmion lattice and the effective magnetic field suggest the possibility of topological moir\'e bands for $\mathcal{H}_{\uparrow}$. However, even in the adiabatic limit of 
long moir\'e periods, the quantum particles experience not
only the $B_{\text{eff}}$ field but also the scalar potential $\Delta_0(\bm r)+|\bm \Delta(\bm r)|-D(\bm r)$.  Here  $D(\bm r)=\left(\hbar^2 / 8 m^*\right) \sum_{i=x, y}\left[\partial_i \boldsymbol{n}\right]^2$, which arises from the precession of the layer-pseudospin \cite{Yu2019Giant,Pan2020Band,Nicolas2024Magic,li2025variational}. 
Because of the interplay among different terms, the topology of moir\'e bands depends on model details. A systematic investigation of band topology as a function of model parameters can be found in Ref.~\cite{Pan2020Band}. A rule of thumb is that the topmost moir\'e valence band has a nonzero Chern number if the low-energy particles are confined to
one layer at $\mathcal{R}_{M}^{X}$ positions and to the other at $\mathcal{R}_{X}^{M}$ positions, forming an effective buckled honeycomb lattice \cite{Pan2020Band}. 

The model parameters can be obtained from first-principles band structure
calculations, which have been carried out for moir\'e superlattices using large-scale density functional theory (DFT) simulations \cite{Devakul2021,Reddy2023Fractional,Wang2024Fractional,Xu2024Maximally,Mao2024Transfer,Zhang2024Polarization,Jia2024Moire,zhang2024universal}. Although the predictions of 
different studies differ in detial, they consistently indicate that the low-energy moir\'e valence bands in both $t$MoTe$_2$ and $t$WSe$_2$ are topological. A representative set of parameters, extracted by fitting the DFT band structure of $t$MoTe$_2$ at a twist angle $\theta = 3.89^\circ$ is: effective mass $m^* \approx 0.6 m_e$, potential amplitude $V = 20.8$ meV, phase $\psi = 107.7^\circ$, and interlayer tunneling strength $w = -23.8$ meV, where $m_e$ is the electron bare mass \cite{Wang2024Fractional}. The monolayer MoTe$_2$ lattice constant is $a_0 = 3.52$ Å.  The skyrmion field $\bm{n}(\bm{r})$ in Fig.~\ref{fig:1}(d) and the corresponding effective magnetic field $B_{\text{eff}}(\bm{r})$ in Fig.\ref{fig:1}(e) were calculated using these parameters with $\theta = 3.8^\circ$. $B_{\text{eff}}(\bm{r})$ has strong spatial variations, reaching peak values of
2400 T for $\theta = 3.8^\circ$—an order of magnitude larger than the average value of 170 T—at the three points in each unit cell that are 
midway between $\mathcal{R}_{M}^{X}$ and $\mathcal{R}_{X}^{M}$ points.

The moir\'e band structure with $\theta = 3.8^\circ$ is shown in Fig.\ref{fig:1}(f). It is calculated using the  Hamiltonian $\mathcal{H}_{\uparrow}$ with the parameters listed above for $t$MoTe$_2$. The first (topmost) moir\'e valence band is energetically isolated from other bands, has a narrow bandwidth of about 8 meV, and carries a finite Chern number of $C_{+K,1}=+1$.  Here we use $C_{\pm K,n}$ to denote the Chern number of the $n$th moir\'e band in the $\pm K$ valley. Due to time-reversal symmetry $\mathcal{T}$, the Chern numbers at opposite valleys are related by $C_{-K,n}=-C_{+K,n}$. In Fig.~\ref{fig:1}(f), the Chern numbers for the second and third bands are, respectively, $-1$ and 0. In this case then the total Chern number of the first two bands is zero, allowing a tight-binding description in terms of two Wannier states, which are polarized to opposite layers and localized, respectively, at $\mathcal{R}_{M}^{X}$ and $\mathcal{R}_{X}^{M}$ positions \cite{Wu2019Topological,Devakul2021,Qiu2023}. As illustrated in Fig.~\ref{fig:1}(g), this tight-binding model, defined on an effective honeycomb lattice, includes complex next-nearest-neighbor hopping with the same pattern as in the Haldane model, reproducing its
nontrivial band topology. Therefore, the first two bands realize the Haldane model \cite{Haldane1988Model} within a single valley and effectively implement the Kane-Mele model (i.e., two time-reversal partners of Haldane model) \cite{Kane2005Quantum} when both valleys are taken into account. The third moir\'e band is topologically trivial and can be described by a tight-binding model based on one Wannier orbital localized at the $\mathcal{R}_{M}^{M}$ position \cite{Qiu2023}.

The Chern numbers of the moir\'e bands vary with the twist angle $\theta$, reflecting the explicit $\theta$-dependence of the moir\'e Hamiltonian through the moir\'e period. In addition, the model parameters $(V, \psi, w)$ also evolve with $\theta$ because of 
lattice relaxation effects. Large-scale DFT calculations for $t$MoTe$_2$ in Refs.~\cite{Zhang2024Polarization,zhang2024universal} reveal the following behavior:  
\begin{equation}
\begin{aligned}
(C_{+K,1}, C_{+K,2}, C_{+K,3}) = 
\begin{cases} 
(1, 1, 1) & \theta = 2.13^\circ \\ 
(1, 1, -2) & 2.45^\circ\leq \theta \leq 2.88^\circ \\ 
(1, -1, 0) & 3.15^\circ \leq \theta \leq 3.89^\circ 
\end{cases}.   
\end{aligned}
\end{equation}
Remarkably, at $\theta = 2.13^\circ$, the first three moir\'e bands carry the same Chern number of $+1$, closely resembling the sequence of Landau levels induced by an external magnetic field.

Similar topological physics emerges in $t$WSe$_2$. Large-scale DFT calculations in Ref.~\cite{zhang2024universal} show that for $2.13^\circ \le \theta \le 3.89^\circ$, the Chern numbers of the first three $+K$ valley moir\'e valence bands in $t$WSe$_2$ are $(C_{+K,1}, C_{+K,2}, C_{+K,3}) = (+1, +1, -1)$.  Another study \cite{Zhang2024Polarization} reports that $(C_{+K,1}, C_{+K,2})$ flips to $(-1,-1)$ in $t$WSe$_2$ when 
$\theta$ decreases to $1.25^{\circ}$.  This transition arises from a reversal of the local layer polarization at $\mathcal{R}_M^{X}$ and $\mathcal{R}_X^{M}$) stacking points driven by a competition between moir\'e ferroelectricity and piezoelectric effects \cite{Zhang2024Polarization}. At a given twist angle the moir\'e bands in $t$WSe$_2$ generally exhibit larger bandwidths than those in $t$MoTe$_2$, which can be attributed to the smaller effective mass $m^*$ in $t$WSe$_2$.

In addition to twist-angle, the out-of-plane electric displacement field provides another 
effective tuning knob to control band topology. The displacement field breaks the layer degeneracy by generating a potential difference between the two layers. As a result, the layer polarization and  Berry curvature distribution of the moir\'e bands can be altered. Topological phase transitions occur at critical displacement field values at which the Chern numbers of specific bands change discontinuously. Such field-induced transitions can be demonstrated using the continuum moir\'e Hamiltonian, which lays the foundation for electrically controllable topological states \cite{Wu2019Topological}. 

The presence of narrow moir\'e bandwidths significantly enhances the importance of Coulomb interactions relative to kinetic energy, often leading to interaction-driven ground states that go beyond single-particle physics. For example, the bandwidth of $t$MoTe$_2$ remains on the order of 10 meV for $\theta$ ranging from $2^{\circ}$ to $4^{\circ}$ \cite{zhang2024universal}, corresponding to a moir\'e period $a_M$ between 10 nm and 5 nm.  The characteristic energy scale of Coulomb interactions is $e^2/(\epsilon a_M)$, which varies from 14 to 28 meV over the same $a_M$ 
range, assuming a dielectric constant $\epsilon = 10$. This interaction energy scale is comparable to, or even larger than, the single-particle bandwidth of the topmost moir\'e valence band, placing the system in the moderate to  strongly correlated regime \cite{Wu2019Topological,Devakul2021,Wang2024Fractional,zhang2024universal}.
The interplay between electron correlations and nontrivial band topology—together with valley (spin), layer, and orbital degrees of freedom leads to a rich landscape of quantum phases that we now discuss.

\begin{figure*}[t]
    \includegraphics[width=1.7\columnwidth]{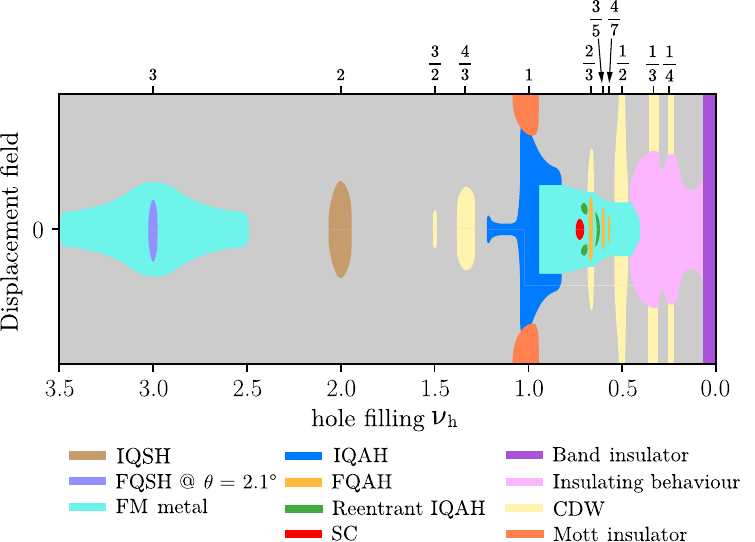}
    \caption{Schematic phase diagram of $t$MoTe$_2$ as a function of hole filling $\nu_h$ and displacement field. This phase diagram summarizes experimental observations in Refs.~\cite{Anderson2023Programming,Cai2023,Zeng2023,Park2023,Xu2023Observation,Park2025Ferromagnetism,Xu2025Interplay,Xu2025signatures,kang2024evidence} across
    twist angles ranging from $2.1^{\circ}$ to $3.9^{\circ}$. Some of the phases appear only at certain devices and twist angles. For example, the superconductor (SC) around $\nu_h \approx 0.73$ has so far been reported only in a device of $t$MoTe$_2$ with $\theta=3.83^{\circ}$ \cite{Xu2025signatures}. Evidence of an FQSH insulator has been reported in $t$MoTe$_2$ with $\theta=2.1^{\circ}$ \cite{kang2024evidence}. (FM metal refers to ferromagnetic metals.)} 
    \label{fig:2}
\end{figure*}

\section{Overview of Quantum Phases}
Recent experimental breakthroughs have revealed a wide variety of quantum phases in $t$MoTe$_2$ and $t$WSe$_2$, including, for example, integer and fractionalized topological states, as well as superconductivity. We begin by providing an overview of the available 
experiments, starting with $t$MoTe$_2$. In Fig.~\ref{fig:2} we present a 
schematic phase diagram versus hole filling factor $\nu_h$ and out-of-plane displacement field.
The phase diagram is based on the experimental observations in Refs.~\onlinecite{Anderson2023Programming,Cai2023,Zeng2023,Park2023,Xu2023Observation,Park2025Ferromagnetism,Xu2025Interplay,Xu2025signatures,kang2024evidence}. Here $\nu_h$ denotes the number of doped holes per moir\'e cell. At $\nu_h = 0$, the system is a bulk intrinsic $t$MoTe$_2$ semiconductor without carrier doping and is therefore a band insulator.

An integer quantum anomalous Hall (IQAH) insulator has been observed at $\nu_h = 1$ around zero displacement field over a broad range of twist angles $\theta$ from $2.1^\circ$ to $3.9^\circ$ \cite{Park2023,Xu2023Observation,Park2025Ferromagnetism,Xu2025Interplay,Xu2025signatures}. This phase is characterized by a Hall resistance $R_{xy}$ with magnitude quantized to $h/e^2$ and a vanishing longitudinal resistance $R_{xx}$ at zero external magnetic field. In the IQAH state at $\nu_h = 1$, holes fully occupy the topmost moir\'e valence band in a single valley, as a result of flatband ferromagnetism driven by Coulomb interactions. 
Spontaneous valley polarization along with nontrivial band topology explains a IQAH phase that breaks time-reversal symmetry \cite{Qiu2023,Li2024Electrically}.
The emergence of ferromagnetism for the magnetic insulator at small displacement fields instead of the antiferromagnetism common in Hubbard-model systems, is 
related to the topologically nontrivial nature of the flat band. The IQAH insulator remains stable up to a critical displacement field, beyond which it undergoes a phase transition to a topologically trivial  Mott insulator with carriers polarized to a single layer.  

Fractional quantum anomalous Hall (FQAH) insulators emerge at fractional fillings $\nu_h=2/3, 3/5$, and $4/7$, displaying vanishing longitudinal resistances $R_{xx}$ and Hall resistances $R_{xy}$ with magnitudes quantized to $h/(\nu_h e^2)$ \cite{Park2023,Xu2023Observation,Park2025Ferromagnetism,Xu2025Interplay,Xu2025signatures}.
Like the IQAH insulator, the FQAH insulators have spontaneous spin/valley 
ferromagnetism persist up to critical displacement fields, after which they undergo a phase transition to a charge density wave (CDW) state at $\nu_h = 2/3$ or metallic states at $\nu_h = 3/5$ and $\nu_h = 4/7$.

In both IQAH and FQAH insulators, the Hall resistances $R_{xy}$ exhibits a hysteresis loop and switches sign when the out-of-plane magnetic field $B$ is swept around zero, as is
typical for spontaneous ferromagnetism \cite{Park2023,Xu2023Observation,Park2025Ferromagnetism,Xu2025Interplay,Xu2025signatures}. Furthermore, the carrier density $n$ in the IQAH and FQAH insulators varies linearly with the $B$ field, as described by the Streda formula \cite{Streda1975Galvanomagnetic}:
\begin{equation}
    \sigma_{xy}=-e \frac{\partial n}{\partial B}=-\mathcal{C}\frac{e^2}{h},
\label{eq:Streda}
\end{equation}
where $\sigma_{xy}$ is the Hall conductivity, $e$ is the elementary charge, and $\mathcal{C}$  is the Chern number of the many-body system. In these IQAH and FQAH insulators, $|\mathcal{C}|$ is quantized to the corresponding $\nu_h$ at zero magnetic field. In the Streda formula, the derivative $\partial n/\partial B$ is taken within the gap of the IQAH (or FQAH) insulators in theory, and along the trajectory with minimal (ideally vanishing) longitudinal resistance $R_{xx}$ in experiment.

The IQAH and FQAH insulators in $t$MoTe$_2$ have been characterized by several experimental probes in addition to transport. Optical measurements 
have played a key role because they sidestep the difficulty of making devices with Ohmic electrical contacts to the TMD bilayers. Reflective magnetic circular dichroism (RMCD) measurements \cite{Anderson2023Programming,Cai2023,Zeng2023} can identify ferromagnetism, as evidenced by remnant RMCD signals at zero external field for $\nu_h \in (0.4,1.2)$. Photoluminescence (PL) spectra have been employed as topological state sensors \cite{Cai2023}. In particular, the spectrally integrated PL intensity has dips around hole fillings of $\nu_h=1, 2/3,$ and $3/5$, and are attributed to reduced trion populations
when the formation of correlated insulating states depletes the holes available for trion formation. Moreover, the dips show linear shifts in carrier densities with an applied magnetic field that match with the Streda formula in Eq.~\eqref{eq:Streda} for the corresponding $\nu_h$. These optical properties provided evidence for IQAH and FQAH insulators in $t$MoTe$_2$ prior to transport studies, which require more stringent device fabrication procedures.  Similar evidence was also provided by local electronic compressibility measurements in which the states at $\nu_h=1$ and $2/3$ were found to be incompressible and shift in density with an applied magnetic field, again consistent with the corresponding Streda formula \cite{Zeng2023}.

The magnetization of $t$MoTe$_2$ has been further characterized using nanoscale superconducting quantum interference devices, which map the magnetic fringe fields \cite{redekop2024direct}. A magnetic signal has been observed over a range of carrier density and displacement field that is consistent with both optical and transport measurements.  Magnetization jumps, reconstructed from the measured magnetic fields, appear at $\nu_h=1$ and $2/3$ that arise from the topological magnetization contributed by equilibrium chiral edge states. The change in magnetization $\delta m$ across the incompressible bulk gap is theoretically given by, 
\begin{equation}
    \delta m = \mathcal{C} \Delta_g / \Phi_0,
    \label{deltam}
\end{equation}
where $\mathcal{C}$ is the Chern number of the system, $\Delta_g$ is the thermodynamic energy gap, and $\Phi_0=h/e$. Equation ~\eqref{deltam} is related to the Streda 
formula since
\begin{equation}
\label{eq:dmdmu}
\begin{aligned}
    \frac{\partial m}{\partial \mu}\Big|_B = \frac{\partial n}{\partial B}\Big|_\mu 
    = \frac{\mathcal{C}}{\Phi_0}.
\end{aligned}
\end{equation}
In Eq.~\eqref{eq:dmdmu}, the first equality follows from a Maxwell relation \cite{macdonald1994introduction} and the second from the Streda formula in Eq.~\eqref{eq:Streda}.  As the chemical potential $\mu$ varies across the bulk energy gap, it changes by $\Delta_g$,  resulting in the magnetization change in Eq.~\eqref{deltam}. The measured $\delta m$ has been used to estimate $\Delta_g$, yielding values of approximately 14 meV at $\nu_h=1$ and  7 meV at $\nu_h=2/3$ in a device with a twist angle $\theta=3.64^{\circ}$  \cite{redekop2024direct}.

Real-space imaging of quantum states in $t$MoTe$_2$ has also been performed using scanning microwave impedance microscopy with sub-100 nm spatial resolution \cite{ji2024local}.  This technique has visualized insulating bulk states and conductive edge states in the IQAH insulator at $\nu_h=1$ and the FQAH insulators at $\nu_h=2/3$ and $3/5$. These observations are consistent with the bulk-boundary correspondence of the topological states and confirm that quantized anomalous Hall conductances are associated with 
conductive edges.

Beyond the IQAH and FQAH insulators, $t$MoTe$_2$ hosts a rich variety of other phases, as shown in Fig.~\ref{fig:2}. The insulating regime at low fillings $\nu_h \lesssim 0.4$ centered on zero displacement field (pink region in Fig.~\ref{fig:2}) is insulating in current devices, possibly due to charge localization from disorder and/or interactions \cite{Park2023,Xu2023Observation,Park2025Ferromagnetism,Xu2025Interplay,Xu2025signatures}.  Contact issues may also contribute to the observed properties in this
low-density regime. Over this range of $\nu_h$ the system becomes conductive above a 
filling-dependent critical displacement field, except at $\nu_h = 1/3$ and $1/4$ \cite{Xu2025signatures}.  The latter two insulating states can be attributed to the interaction-driven CDW phases expected in triangular lattice extended Hubbard models at 
strong interactions and commensurate fillings and realized in the layer polarized regime under high displacement fields.

Bulk states become metallic for $0.4 \lesssim \nu_h < 1$, except at certain fillings such as $\nu_h = 2/3$, $3/5$, and $4/7$, where FQAH insulators emerge. The metallic states exhibit anomalous Hall effects around zero displacement field (cyan region in Fig.~\ref{fig:2}), indicating itinerant electron ferromagnetism from spontaneous valley polarization \cite{Park2023}. The anomalous Hall metal at $\nu_h=1/2$ has attracted particular attention. Near this filling, the Hall resistance $R_{xy}$  varies linearly with $\nu_h$ and crosses $2e^2/h$ at $\nu_h=1/2$, while the longitudinal resistance $R_{xx}$ remains at a few k$\Omega$, consistent with a compressible state \cite{Park2023}. This behavior resembles that of the composite Fermi liquid near half-filling of the lowest Landau level in high magnetic fields \cite{Willett1987Observation,Jain1989Composite,Halperin1993Theory,Willett1993Experimental,Kang1993How,Goldman1994Detection}, consistent with numerical predictions of a zero-field composite Fermi liquid at $\nu_h = 1/2$ in $t$MoTe$_2$ \cite{Dong2023,Goldman2023Zero}. Optical signatures of this zero-field composite Fermi liquid have also been reported \cite{Anderson2024Trion}.

\begin{figure*}[t]
    \includegraphics[width=2.\columnwidth]{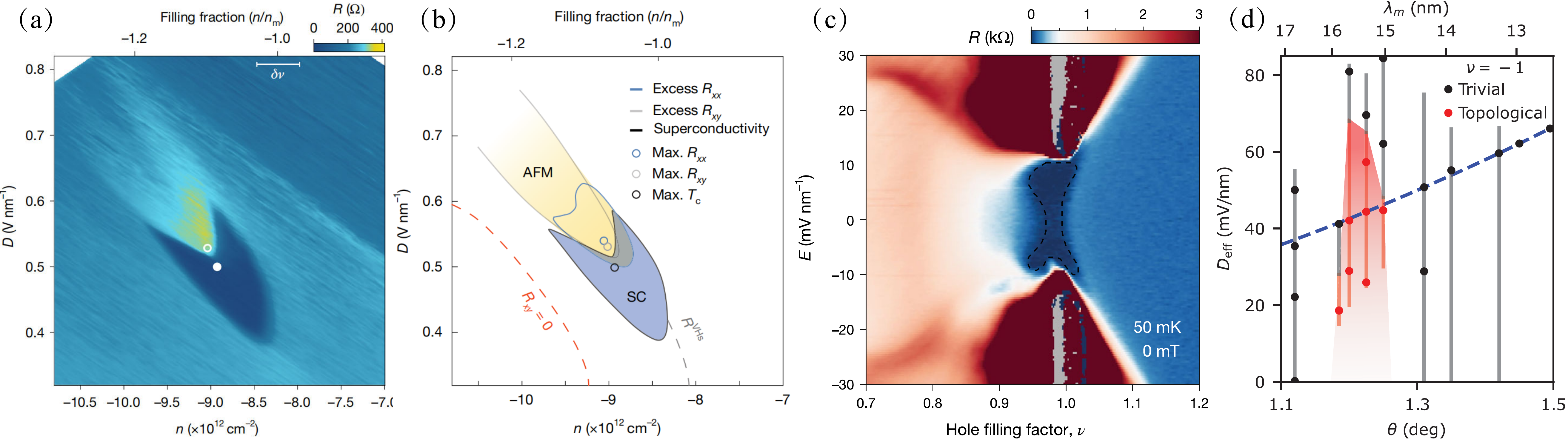}
    \caption{Superconductivity and topological phases observed in $t$WSe$_2$. (a) A map of longitudinal resistance $R$ is plotted as a function of density and displacement field at $T=200\,\mathrm{mK}$ in  $t$WSe$_2$ at $\theta=5.0^\circ$. (b) A schematic phase diagram of superconductor (SC) and antiferromagnetic (AFM) phases based on (a). Both (a) and (b) are adapted from Ref.~\cite{Guo2025Superconductivity}. (c) Experimental observation of superconductivity in $t$WSe$_2$ at $\theta=3.65^\circ$, reported by Ref.~\cite{Xia2025Superconductivity}. The plot shows the longitudinal resistance $R$ as a function of hole filling factor and applied electrical field  with temperature $T=50\,\mathrm{mK}$. The region with zero resistance is marked by a dashed line. (d) Experimental phase diagram of the IQAH insulator at $\nu_h = 1$ as function of $\theta$ and effective displacement field $D_{\text{eff}}$ from Ref.~\cite{Foutty2024Mapping}. Red regions mark the IQAH phase, while black regions represent the topologically trivial phase. 
    } 
    \label{fig:3}
\end{figure*}

A recent transport study performed on a $\theta=3.83^{\circ}$ sample with reduced disorder and enhanced mobility, uncovered additional exotic states at $\nu_h < 1$ \cite{Xu2025signatures}. A reentrant IQAH insulator emerges between the $\nu_h = 2/3$ and $3/5$ FQAH insulators near zero displacement field  \cite{Xu2025signatures}. This phase exhibits a Hall resistance $R_{xy}$ quantized to $h/e^2$ and vanishing $R_{xx}$—hallmarks of the IQAH effect—yet occurs at an incommensurate filling $\nu_h \approx 0.63$, that is not associated with a simple rational fraction. This reentrant IQAH state is reminiscent of the reentrant integer quantum Hall effect seen in Landau levels with short-range disorder \cite{Li2010Observation} and is likely an anomalous Hall crystal with spontaneous translational symmetry breaking. Another reentrant IQAH insulator appears at $\nu_h \approx 0.7$, but only at a finite displacement field. 

Remarkably, signatures of superconductivity have been observed in the same device around zero displacement field for $\nu_h \in (0.71,0.76)$.  The superconducting state
is separated from the $\nu_h=2/3$ FQAH insulator and the reentrant IQAH insulator
by a resistance peak \cite{Xu2025signatures}. The superconducting zero-resistance state is reached up to 300 mK, with a perpendicular critical magnetic field as high as approximately 0.6 T.  Above the superconducting transition temperature, the system enters a metallic phase with an anomalous Hall effect and magnetic hysteresis, indicating that the superconductor likely emerges from a ferromagnetic parent state. 
The embedding of a superconducting phase within a ferromagnetic metal is quite unusual. Moreover, the proximity of this state to the FQAH and reentrant IQAH insulators highlights its unconventional nature and calls for further experimental and theoretical exploration.

For $\nu_h > 1$, correlated insulators have been observed at $\nu_h = 3/2$ and $4/3$, likely corresponding to CDW states stabilized by electron-electron interactions \cite{Xu2025signatures}. At $\nu_h = 2$, multiple signatures point to an integer quantum spin Hall (IQSH) insulator around zero displacement field \cite{Xu2025Interplay,kang2024evidence}, including: (1) vanishing Hall resistance and the absence of a remnant RMCD signal; (2) quantized two-terminal resistance plateaus at $h/2e^2$; (3) nonlocal transport behavior; and (4) strong magnetoresistance under a small in-plane magnetic field combined with weak response to an out-of-plane field. These observations suggest that the $\nu_h = 2$ state is nonmagnetic and realizes the quantum spin Hall insulator expected \cite{Wu2019Topological} in the absence of interactions.  In this state the topmost moir\'e valence bands from the $\pm K$ valleys—characterized by opposite out-of-plane spins $S_z$ and opposite Chern numbers—are fully hole doped. Helical edge states have opposite velocities for opposite $S_z$. In TMD homobilayer moir\'e materials the Ising $S_z$  symmetry provides
additional protection for the helical edge states on top of that afforded by time reversal symmetry. An in-plane (out-of-plane) magnetic field breaks (preserves) the  $S_z$ symmetry, explaining the magnetoresistance behavior.

For $2<\nu_h<4$, the second moir\'e valence bands of the $\pm K$ valleys are partially filled. Ferromagnetism has been identified for $\nu_h \in (2.6,3.7)$ 
by observing both the quantum anomalous Hall effect and a remnant RMCD signal \cite{Park2025Ferromagnetism,Xu2025Interplay,An2025Observation}. The ferromagnetic metal at $\nu_h=3$ can be converted to a Chern insulator with an integer 
quantized Hall effect by applying an out-of-plane magnetic field \cite{Park2025Ferromagnetism,Xu2025Interplay}.  Interestingly, for a given direction of the magnetic field, the Chern number of the incipient Chern insulator at $\nu_h = 3$ has the opposite sign to that of the IQAH insulator at $\nu_h = 1$ for $\theta \geq 3.1^{\circ}$, but shares the same sign at $\theta = 2.6^{\circ}$. This observation supports the notion
that the Chern number of the second moir\'e band is twist-angle–dependent
suggested by large-scale DFT calculations \cite{Zhang2024Polarization}.

A transport study of $t$MoTe$_2$ with $\theta=2.1^{\circ}$ revealed an IQAH insulator at $\nu_h=1$, and IQSH insulators at $\nu_h=2, 4, $ and $6$ with two-terminal conductancea of $2e^2/h$, $4e^2/h$, and  $6e^2/h$, respectively \cite{kang2024evidence}.  
The observation of these IQSH insulators is consistent with large-scale DFT calculations \cite{Zhang2024Polarization}, which show the first three moir\'e valence bands in $+K$ ($-K$) valleys all carry the same Chern number of $+1$ ($-1$) for $\theta$ around $\theta=2.1^{\circ}$. The multiple pairs of edge states at $\nu_h=4$ and $6$ are protected by the Ising $S_z$ symmetry. 
Moreover, observations suggestive of a fractional quantum spin Hall (FQSH) insulator has been obtained at the odd filling factor $\nu_h = 3$, where the two-terminal conductance reaches $3e^2/h$, with each edge contributing a fractional conductance of $3e^2/2h$. This state has been shown to be incompressible with pronounced nonlocal resistance  properties \cite{kang2024evidence}.
A proposed theoretical scenario attributes these observations to a 
FQSH effect at half-filling of the second moir\'e valence bands in both valleys 
in which Coulomb interactions induce correlated insulating states with fractionalized helical edge modes \cite{kang2024evidence}. It should be noted that time-reversal symmetry breaking has been observed in this state through a weak anomalous Hall effect\cite{kang2025time} that is destroyed
by an extremely tiny magnetic field $\sim 20$ mT. Independently, Ref.~\cite{li2025universal} reported spontaneous ferromagnetism at $\nu_h=3$ for $\theta$ around $2.1^{\circ}$ using RMCD measurement and scanning
nanoSQUID-on-tip  magnetometry.  The nanoSQUID magnetometry measurements 
did not detect an orbital magnetization jump, which is expected at topological gaps. The spatially resolved measurement in Ref.~\cite{li2025universal} suggests
that the $\nu_h=3$ correlated topological phases could be obscured
by the device disorder. Considering these observations altogether, the nature of the $\nu_h=3$ state around $\theta \approx 2.1^{\circ}$ remains unsettled.

We now turn to describe quantum phases in $t$WSe$_2$, which is a close cousin of $t$MoTe$_2$ but exhibits qualitatively different physics. A pioneering transport study ~\cite{Wang2020Correlated} of $t$WSe$_2$  uncovered a correlated insulating state at $\nu_h=1$ across twist angle $\theta \in (4^{\circ},5.1^{\circ})$. This correlated insulator is tunable by the applied displacement field, with the largest insulating gap appearing at a finite, $\theta$-dependent field strength. The insulator does not exhibit the quantum anomalous Hall effect, and is therefore, topologically trivial. In this range of twist angles, the bandwidth of the first moir\'e valence band in $t$WSe$_2$ is sizable, reaching approximately 100 meV at $\theta = 5.08^\circ$ according to DFT calculations. The corresponding Coulomb interaction energy scale $e^2/(\epsilon a_M)$ is about 40 meV at $\theta = 5.08^\circ$, assuming $\epsilon=10$, which is comparable to but smaller than the bandwidth. In this regime of moderate correlations, the Coulomb interaction is insufficient to induce valley polarization via Stoner ferromagnetism but can stabilize intervalley-coherent antiferromagnetic states that are topologically 
trivial correlated insulators at $\nu_h = 1$.  The $\nu_h=1$ insulating states 
seem to appear when the Van Hove singularities in the single-particle moir\'e band structure, which evolve with the displacement field, are close to the Fermi energy. When these singularities pass through $\nu_h = 1$, the diverging density of states enhances correlation effects and strengthens the insulating gap. This mechanism underlies the observed field tunability of the correlated insulator. Moreover, at $\theta=5.1^{\circ}$, zero-resistance pockets were observed on doping away from $\nu_h=1$, which 
appeared to indicate a transition to a superconducting state~\cite{Wang2020Correlated}.

Robust demonstrations of superconductivity in $t$WSe$_2$ have been achieved recently. One experiment reported superconductivity in $t$WSe$_2$ at a twist angle of $5.0^{\circ}$, with a maximum critical temperature of 426 mK \cite{Guo2025Superconductivity}. The superconducting phase emerges in a narrow region at finite displacement fields near $\nu_h = 1$, closely tracking the van Hove singularities, as shown in Figs.~\ref{fig:3}(a) and \ref{fig:3}(b). Adjacent to this phase is a metallic state exhibiting Fermi surface reconstruction, likely driven by intervalley coherent antiferromagnetic order. A sharp phase boundary separates the superconducting and metallic states at low temperatures. A separate experiment reported superconductivity in $t$WSe$_2$ at smaller twist angles of $3.5^{\circ}$ and $3.65^{\circ}$ \cite{Xia2025Superconductivity}. In these systems, superconductivity also emerges near $\nu_h = 1$ but around zero displacement field and away from the van Hove singularities, as illustrated in Fig.~\ref{fig:3}(c). The optimal superconducting transition temperature is about 200 mK.  The superconductor borders on metallic states below and above $\nu_h=1$ at zero displacement field, but undergoes a continuous transition to a correlated insulator by increasing the displacement field.  These results demonstrate a rich and tunable interplay between electronic band structure, superconductivity, and correlated states in $t$WSe$_2$, highlighting phenomena that remain to be fully understood and explored.

Topological states have been observed in $t$WSe$_2$ at smaller twist angles. Local electronic compressibility measurements of $t$WSe$_2$ have been performed using scanning single-electron transistor microscopy at small $\theta$ from $1.1^{\circ}$ to $1.5^{\circ}$.\cite{Foutty2024Mapping} Around $\theta=1.23^{\circ}$, incompressible states were identified at $\nu_h=1$ and $3$, which shift in density with an applied out-of-plane magnetic field following the Streda formula in Eq.~\eqref{eq:Streda}. The persistence of these incompressible states down to zero magnetic field indicates the presence of IQAH insulators. The field dependence of their trajectories reveals that the IQHA states 
at $\nu_h = 1$ and $3$ carry the same Chern number with a magnitude of 1. An experimental phase diagram for  $\nu_h = 1$ as a function of twist angle and tip-induced displacement field is shown in Fig.~\ref{fig:3}(d). While the suppression of the IQAH phase by a finite displacement field is consistent with theoretical expectations, the emergence of the IQAH insulator within a narrow twist angle window centered at $1.23^{\circ}$ is striking. Outside this window in Fig.~\ref{fig:3}(d), Chern insulators can still be stabilized by an external magnetic field.

In a separate experiment combining magnetic circular dichroism and exciton sensing techniques, Chern insulators near $\nu_h = 1$ were also observed in $t$WSe$_2$ at twist angles ranging from $1.8^{\circ}$ to $2.7^{\circ}$ under an applied magnetic field \cite{Knuppel2025Correlated}. These findings are consistent with the fact that the moir\'e bands become progressively narrower at smaller twist angles, enhancing the tendency toward flatband ferromagnetism and spontaneous valley polarization. This provides an explanation for the emergence of Chern insulators in $t$WSe$_2$ at small twist angles, either at zero or finite magnetic field. 

IQSH effects have also been reported in $t$WSe$_2$ with $\theta=3.0^{\circ}$ at even filling factors of $\nu_h=2$ and $4$ \cite{Kang2024Double}. The observed signatures include: (1) quantized (within 10\%) resistance plateaus of $h/(\nu_h e^2)$; (2) large nonlocal transport signals; (3) an insulating bulk; and (4) resistance that is insensitive to out-of-plane magnetic fields but increases under an in-plane magnetic field. These signatures are consistent with those observed in the IQSH insulators of $t$MoTe$_2$ and align with quantum transport of helical edge states in a system with spin Chern bands protected by Ising $S_z$ symmetry.

\section{Integer Topological States}

Building on the preceding overview of quantum phases in $t$MoTe$_2$ and $t$WSe$_2$, we now delve deeper into first integer and then fractional topological states, as well as superconductivity. In particular, we compare these phases to their counterparts in related systems and discuss the theoretical mechanisms underlying their emergence.

We begin with integer topological states, specifically the IQAH and IQSH insulators, which are characterized by integer-valued topological invariants. The discovery of the integer quantum Hall effect \cite{Klitzing1980New} in two-dimensional electron systems under strong magnetic fields marked the beginning of the study of topological electronic states in condensed matter physics. The quantization of Hall conductance in this effect is understood in terms of the Chern number $\mathcal{C}$ \cite{Thouless1982Quantized,Niu1985Quantized}, which determines the Hall conductivity in units of $e^2/h$ and corresponds to the net chirality of the edge modes. Quantized Hall transport in finite-size devices can 
normally be understood in terms of currents carried by chiral edge states, while the bulk state is insulating.   The chiral edge states are protected from backscattering and this property underlies the remarkable robustness of the quantum Hall effect.

The IQAH insulator is a generalization of the integer quantum Hall insulator \cite{Haldane1988Model,Onoda2003Quantized,Liu2008Quantum,Yu2010Quantized}, and is likewise characterized by a nonzero Chern number, chiral edge states, and quantized Hall conductance. 
However the IQAH insulator appears in the absence of an external magnetic field and 
therefore requires spontaneous breaking of time-reversal symmetry. 
IQAH insulators have been realized in magnetically doped topological insulator thin films \cite{Chang2013Experimental}, few-layer MnBi$_2$Te$_4$ \cite{Zhao2020Tuning,Deng2020Quantum,Liu2020Robust,Zhang2019Topological,Li2019Intrinsic,Otrokov2019Prediction}, graphene-based moir\'e systems \cite{Sharpe2019Emergent,Serlin2020Intrinsic,Chen2020Tunable,Polshyn2020Electrical,Polshyn2022Topological}, and TMD-based moir\'e homobilayers and heterobilayers - specifically MoTe$_2$/WSe$_2$ \cite{Li2021}. 

\begin{figure}[t]
    \includegraphics[width=1.\columnwidth]{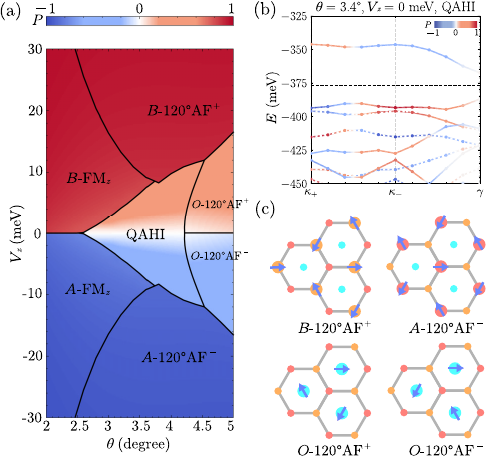}
   \caption{Competing topological and magnetic phases at $\nu_h = 1$.  (a) Quantum phase diagram at $\nu_h = 1$ as a function of displacement potential $V_z$ and twist angle $\theta$. The color map represents the layer polarization $P$. (b) The $\nu_h = 1$ mean-field band structure at $\theta = 3.4^\circ$ and $V_z=0\,\mathrm{meV}$ for the integer quantum anomalous Hall insulator (QAHI). The solid (dashed) lines plot bands in
   the $+K$ ($-K$) valley, and the horizontal black dashed line indicates the chemical potential centered in the interaction-induced gap. The color denotes the layer polarization of Bloch states, which varies dramatically across the moir\'e Brillouin zone at energies above the chemical potential, indicating the winding of layer pseudospin.  (c) Schematic illustration of the four $120^\circ$ antiferromagnetic states in (a) with different spatial occupations and spin vector chiralities. The red, orange, and cyan dots represent the $\mathcal{R}_M^X$, $\mathcal{R}_X^M$, and $\mathcal{R}_M^M$ sites in the moir\'e superlattices. $\text{AF}^{\pm}$ denote antiferromagnetic states with $\pm$ spin vector chirality. This figure is adapted from Ref.~\cite{Li2024Electrically}.} 
   \label{fig:4}
\end{figure}

In twisted TMD homobilayers, the valley-polarized IQAH insulator competes with other 
broken-symmetry states at $\nu_h=1$. This is most evident in $t$WSe$_2$, where the 
correlated insulators observed at $\nu_h=1$ for $\theta$ above about $3^{\circ}$ \cite{Wang2020Correlated,Guo2025Superconductivity,Xia2025Superconductivity} are topologically trivial, although the  underlying single-particle moir\'e bands carry valley-contrasting Chern numbers \cite{zhang2024universal}.  This correlated insulator is normally interpreted as a valley-coherent state with a finite ordering wave vector that 
is expected to result in $120^{\circ}$ antiferromagnetism in real space \cite{Pan2020Band, Qiu2023}.  

Figure~\ref{fig:4}(a) presents a theoretical phase diagram of $t$MoTe$_2$ at $\nu_h = 1$, obtained within the mean-field Hartree-Fock approximation \cite{Li2024Electrically}, as a function of twist angle $\theta$ and the interlayer potential difference $V_z$ induced by an out-of-plane displacement field. At $V_z = 0$, the phase diagram includes three distinct phases: an IQAH insulator at intermediate $\theta$ (with its mean-field band structure shown in Fig.~\ref{fig:4}(b)), an intervalley-coherent antiferromagnetic state at large $\theta$, and a layer-polarized ferromagnetic state at small $\theta$. At large $\theta$, weaker correlation effects favor the formation of valley-coherent states over valley-polarized ones.  In contrast, at very small $\theta$, strong Coulomb interactions drive charge localization onto either $\mathcal{R}_M^X$ or $\mathcal{R}_X^M$ sites, leading to spontaneous layer polarization. The phase boundaries are also sensitive to $V_z$: when $V_z$ exceeds critical values, the IQAH phase is replaced by a layer-polarized, topologically trivial correlated insulators similar to those that appear in 
extended Hubbard models on triangular lattices, which can be independently realized in moir\'e TMD heterobilayers \cite{Wu2018Hubbard,Regan2020Mott,Tang2020Simulation}. 
We note that this mean-field phase diagram is qualitative, since it neglects effects like lattice relaxation and beyond-mean-field correlations. Experimental signatures of the competition between the valley polarized states and other magnetic states in $t$MoTe$_2$ have been reported in Ref.~\cite{chang2025evidence}. Because the topological and magnetic phases shown in Fig.~\ref{fig:4} are symmetry-breaking phases driven by interactions, they host a variety of low-energy collective excitations, such as excitons, magnons, and domain walls, which are currently under active investigation \cite{Qiu2025Quantum, Qiu2025Topological, wang2024diverse, zhou2025itinerant,xiong2025propagating}.

The IQSH insulator is a generalization of the IQAH insulator \cite{Kane2005Quantum,Bernevig2006Quantum,Sheng2006Quantum}, characterized by topological bands with spin-contrasting Chern numbers. The IQSH state is bulk insulating and, when protected by Ising spin symmetry, can host multiple helical edge channels. If Ising spin symmetry is broken but time-reversal symmetry is preserved, the IQSH can become a time-reversal invariant two-dimensional topological insulator, classified by a $Z_2$ topological invariant \cite{Kane2005Z2}. In this case, time-reversal symmetry protects a single helical mode per edge, consisting of a pair of counterpropagating edge states with opposite spins. For symmetry-preserving disorder, the helical modes are also immune to backscattering. Therefore, each helical mode contributes $e^2/h$ to the two-terminal conductance. Similar transport signatures of two-dimensional topological insulators have been previously reported in HgTe quantum wells \cite{Konig2007Quantum}, InAs/GaSb quantum wells \cite{Knez2010Finite,Knez2011Evidence}, and monolayer WTe$_2$ \cite{Fei2017Edge,Wu2018Observation}.

In twisted TMD homobilayers, the total spin $S_z$ is approximately conserved and can serve as an emergent effective symmetry. 
The IQSH insulator at $\nu_h = 2$ exhibits a two-terminal conductance of $2e^2/h$, protected by both $S_z$ and time-reversal symmetry. In contrast, the IQSH insulators at $\nu_h = 4$ and $6$ exhibit two-terminal conductance values of $\nu_h e^2/h$, arising from multiple helical edge modes protected solely by the approximate $S_z$ symmetry. We also note that other competing states may exist at $\nu_h = 2$ \cite{Qiu2023,Luo2024,Liu2024Gate}, but experimental evidence for these is so far lacking.

\section{Fractionalized Topological States}
Two-dimensional electron systems under strong magnetic fields exhibit not only the integer quantum Hall effect but also the fractional quantum Hall effect \cite{Tsui1982Two}, a paradigmatic example of a strongly correlated topological phase. At certain fractional Landau level fillings, electron interactions stabilize incompressible quantum liquids with fractionally quantized Hall conductance. A prototypical example is the fractional quantum Hall insulator (FQHI) at filling factor $1/3$, which exhibits a Hall conductivity of $(1/3)e^2/h$ and is well described by the Laughlin wavefunction \cite{Laughlin1983}. FQHIs occur at various rational fillings with odd denominators — such as $1/3$, $2/5$, and $3/7$ — and less commonly at even-denominator fractions like $5/2$. These phases host quasiparticle excitations with fractional charge and anyonic statistics, as exemplified by the Laughlin quasiparticles and quasiholes in the $1/3$ state. 

The concept of FQHIs was first extended to the zero-magnetic-field case 
in a series of theoretical studies of model lattice systems  \cite{Tang2011,Sun2011,Neupert2011,Regnault2011Fractional,Sheng2011Fractional}. 
The zero-field states were referred to as fractional Chern insulators (FCIs).
These studies focused on systems with partially filled topological flat bands with nonzero Chern numbers that emulate Landau levels, allowing strong electron interactions to stabilize fractionalized phases analogous to those in the FQHI. FCIs exhibit key hallmarks of fractional quantum Hall physics —including fractionally quantized Hall conductance, anyonic excitations, and topological ground-state degeneracy — but can also display new features, going beyond the physics of Landau levels \cite{Bergholtz2013topological}. 

\begin{figure*}[t]
    \includegraphics[width=2.\columnwidth]{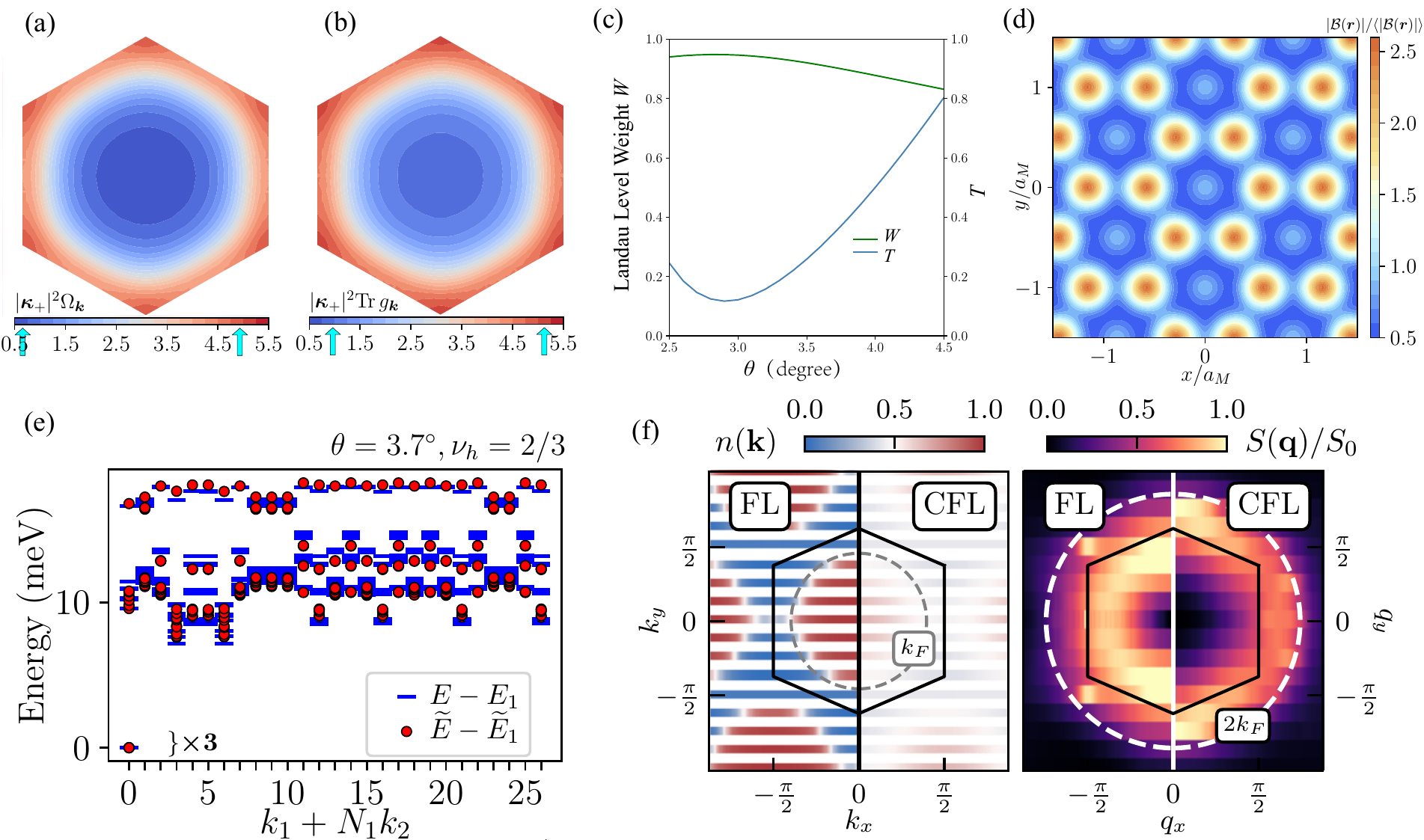}
    \caption{ Numerical results in the first moir\'e valence band of $t$MoTe$_2$ for fractionalized states. (a-b) Berry curvature $\Omega_{\bl k}$ and trace of quantum metric $\mathrm{Tr} g_{\bl k}$ at $\theta = 2.9^\circ$ in the moir\'e Brillouin zone. (c) Deviation $T$ from ideal quantum geometry and Laudau level weight $W$ as functions of $\theta$. (d) Map of $|\mathcal B(\bl r)|$ scaled by its spatial average at $\theta = 2.9^\circ$. (e) ED spectra based on $\varphi_{1,\boldsymbol{k}}$  (blue lines) and  $\Theta_{\boldsymbol k}(\boldsymbol r)$ (red dots) at $\nu_h = 2/3$ and $\theta = 3.7^\circ$. (f) Infinite DMRG results of the composite Fermi liquid (CFL) at $\nu_h = 1/2$ and $\theta = 3.7^\circ$. The left panel shows the occupations $n(\bl k)$ in the Brillouin zone, comparing the Fermi liquid (left side) and the CFL (right side). The right panel presents the connected structure factor $S(\bl q) = \langle \hat{\rho}_{\bl q} \hat{\rho}_{-\bl q} \rangle - \langle \hat{\rho}_{\bl q} \rangle \langle \hat{\rho}_{-\bl q} \rangle$. Panels (a)-(e) are adapted from Ref.~\cite{li2025variational}, and panel (f) is from Ref.~\cite{Dong2023}.  
    } 
    \label{fig:5}
\end{figure*}

The definition of FCIs was later expanded to include lattice-like experimental systems
at finite magnetic fields.  FCIs have been observed in several van der Waals heterostructures. Magnetocapacitance measurements have provided evidence of FCIs at fractional fillings of Harper-Hofstadter bands, arising from the interplay of a strong magnetic field ($\sim$30 T) and a superlattice potential in a bilayer graphene–hexagonal boron nitride heterostructure \cite{Spanton2018Observation}. In magic-angle twisted bilayer graphene, FCIs have been reported at lower magnetic fields ($\sim$5 T) through high-resolution local compressibility measurements \cite{Xie2021Fractional}. FCIs at zero magnetic field, known as FQAH insulators, were first realized in $t$MoTe$_2$ \cite{Cai2023,Zeng2023,Park2023,Xu2023Observation}, a phenomenon now well established through various experimental techniques. FQAH insulators have also been observed in another moir\'e system, multilayer rhombohedral graphene aligned with hexagonal boron nitride \cite{lu2024fractional,Xie2025Tunable}.

The emergence of FQAH insulators within a valley-polarized Chern band in $t$MoTe$_2$ can be theoretically understood by mapping the band to Landau levels, a process that can be achieved through two complementary approaches. One approach utilizes the adiabatic approximation, which incorporates the emergent magnetic field $B_{\text{eff}}$ generated by the skyrmion texture, shown in Eq.~\eqref{eq:bz}, in the moir\'e Hamiltonian \cite{Nicolas2024Magic,shi2024adiabatic,li2025variational}. A related approach is based on the observation that the first moir\'e band in $t$MoTe$_2$ has nearly ideal quantum geometry,
in which the Berry curvature $\Omega_{\bm k}$ and the trace of quantum metric $\text{Tr} g_{\bm k}$ fluctuate in sync in momentum space as illustrated in Figs.~\ref{fig:5}(a) and \ref{fig:5}(b). A measure of the deviation from ideal quantum geometry is $T=\frac{1}{2\pi}\int d^2\boldsymbol k\mathrm{Tr}\,g_{\boldsymbol{k}}-|\mathcal{C}|$, where momentum integration is over the moir\'e Brillouin zone and $\mathcal{C}$ is the Chern number and 
$T$ can be shown to be non-negative. A band with ideal quantum geometry has $T=0$ and can be shown to have generalized Landau level Bloch wave functions~\cite{Ledwith2020,Wang2021Exact} of the form
\begin{equation}
    \Theta_{\boldsymbol  k}(\boldsymbol r)=\mathcal N_{\boldsymbol  k}\mathcal{B}(\boldsymbol r)\Psi_{0,\boldsymbol k}(\boldsymbol r),
\end{equation}
where  $\Psi_{0,\boldsymbol k} (\boldsymbol{r})$ is the magnetic Bloch wave function of the zeroth Landau level, $\mathcal{B}(\boldsymbol r)$ is spatially dependent but wavevector $\bm k$-independent, and  $\mathcal{N}_{\boldsymbol{k}}$ is a normalization factor. In $t$MoTe$_2$, $T$ is not zero but can be as small as $0.1$, as shown in Fig.~\ref{fig:5}(c). The wave function $\varphi_{1,\boldsymbol k}$ of the first moir\'e valence band in $t$MoTe$_2$ can be approximated by $\Theta_{\boldsymbol  k}$. A variational approach, developed in Ref.~\cite{li2025variational},  is used to obtain the optimal $\mathcal B(\boldsymbol r)$  by maximizing the Landau level weigth $W$ defined as the momentum average of $|\langle \psi_{\bm{k}} | \Theta_{\boldsymbol k} \rangle|^2$.  The maximized weight $W$ is shown in Fig.~\ref{fig:5}(c), which reaches up to about 0.95. The obtained $\mathcal B(\boldsymbol r)$ function is plotted in Fig.~\ref{fig:5}(d) and has a spacial 
variation that matches that of the total electron density of the first moir\'e band.

The ideal Chern band with the wave function $\Theta_{\boldsymbol k}(\boldsymbol r)$ possesses a distinctive feature: it enables the construction of trial wavefunctions for fractional Chern insulators (FCIs) \cite{Wang2021Exact,Ledwith2020}. These trial states take the form $\Phi_F=\Psi_F\prod_{i} \mathcal{B}(\boldsymbol{r}_i)$, where $\boldsymbol{r}_i$ denotes the coordinate of the $i$th electron, and $\Psi_F$ describes fractional quantum Hall states (or composite Fermi liquid states) within the zeroth Landau level. These constructed wavefunctions $\Phi_F$ are exact ground states for specific short-range repulsive Hamiltonians \cite{Wang2021Exact}. The substantial overlap between the single-particle states $\varphi_{1,\boldsymbol{k}}$ and the generalized Landau level functions $\Theta_{\boldsymbol k}(\boldsymbol r)$ offers a simple physical picture that 
explains the emergence of FQAH insulators in $t$MoTe$_2$. Figure~\ref{fig:5}(e) presents the exact diagonalization (ED) spectra  at $\nu_h=2/3$ based on, respectively, $\varphi_{1,\boldsymbol{k}}$ and  $\Theta_{\boldsymbol k}(\boldsymbol r)$ using Coulomb interaction, which agree quantitatively and clearly reveal an FCI state.

Numerical studies are essential for investigating FCIs. While ED studies are limited by small system sizes and few-band projections, they have proven successful in identifying FCIs through analysis of the energy spectra and quasi-degeneracy of ground states, spectral flow under flux insertion, and the unique counting of the entanglement spectrum below the gap \cite{Li2021Spontaneous,Crepel2023,Nicolas2023Pressure,Goldman2023Zero,Dong2023,Reddy2023Toward,Reddy2023Fractional,Wang2024Fractional,Xu2024Maximally,Jia2024Moire,yu2024fractional,li2025variational}. The density matrix renormalization group (DMRG) method has also been applied to study fractionalized states in $t$MoTe$_2$ \cite{Dong2023,chen2025fractional,He2025Fractional}. For example, numerical evidence of the composite Fermi liquid at $\nu_h=1/2$ in $t$MoTe$_2$ from DMRG calculations is shown in Fig.~\ref{fig:5}(f) \cite{Dong2023}. Recently, a deep learning quantum Monte Carlo method based on neural networks has been developed to study both the integer and fractional topological states in $t$MoTe$_2$ \cite{li2025deep,luo2025solving}.

The presence of spin-contrasting Chern bands in $t$MoTe$_2$ yields physics that goes beyond the Landau level framework, with the experimental signature of the FQSH insulator at $\nu_h = 3$ serving as a prominent example \cite{kang2024evidence}. This has prompted several theoretical studies to explore the nature of this state \cite{zhang2024nonabelian,jian2024minimal,maymann2024theory,villadiego2024halperin,Abouelkomsan2025nonabelian}, with candidate states such as nonabelian spin Hall insulators  and Halperin states. DFT band structure calculations \cite{Zhang2024Polarization,wang2025higher,xu2025multiple, zhang2024universal} have shown that the
second moir\'e band in $t$MoTe$_2$ at $\theta$ around $2^{\circ}$ can mimic the first Landau level.  Numerical evidences of nonabelian fractionalized states at half filling of the second moir\'e band with spontaneous valley polarization have been reported in $t$MoTe$_2$ \cite{wang2025higher,xu2025multiple,Reddy2024nonabelian,Ahn2024nonabelian,Chen2025Robust}. These findings suggest the possibility, not yet realized, that $t$MoTe$_2$ might be a platform for exploring nonabelian topological phases. 

The low-energy collective modes of insulators can 
play a helpful role in characterizing phases of matter, especially at long-wavelengths 
where they couple directly to light.  Numerical calculations have established \cite{shen2024magnetorotons,paul2025shining,hu2024hyperdeterminants} that the 
FCI states of $t$MoTe$_2$ have low-energy intraband collective excitations that are 
similar in many respects to the magnetoroton excitations of conventional fractional quantum Hall
systems.  In the FQHI case, magnetoroton excitations \cite{girvin1986magneto}
are optically dark, {\it i.e.} they do not contribute to the optical conductivity.
In the FCI case it is easy to see \cite{wolf2025intraband} that they are dark in the case of perfectly flat bands since the projection of the current operator to the partially occupied band is zero.  It has been argued that the coupling to photons is anomalously weak \cite{wolf2025intraband}
even when the bands are not perfectly flat, but it can be strengthened \cite{wu2016moire,kousa2025theory} by adding periodic modulation at length scales that are 
longer than the moir\'e period.  Even if optical signals are weak, they might still be 
observable and highly informative.  Collective mode studies of FCIs therefore have considerable promise,
but require progress in making larger devices, improving the energy resolution of near-field optical probes, and/or advancing techniques for THz spectroscopy studies of small devices \cite{de2025roadmap}.

\section{Superconductors}
The observation of superconductivity in $t$WSe$_2$ across multiple devices and twist angles establishes twisted homobilayer TMDs as a second moir\'e platform, after twisted 
graphene bilayer and multilayers, that supports robust superconductivity \cite{Guo2025Superconductivity,Xia2025Superconductivity}. Available experiments show that $t$WSe$_2$ can host superconductivity at $\theta=3.5^{\circ}$, $3.65^{\circ}$, and $5^{\circ}$, suggesting that superconductivity in this system is less sensitive to twist angle compared to graphene-based moir\'e systems. This behavior reflects a general trend in twisted TMD homobilayers, where many-body correlation effects persist over a wider range of twist angles due to a weaker dependence of the moir\'e bandwidth on $\theta$. Because correlations are weak away from the narrow {\em magic} twist-angle regime, it is more difficult to precisely reproduce behavior from one device to another in 
the graphene multilayer case. Nevertheless, it is important to recognize that significant differences exist between $t$WSe$_2$ near $\theta = 5^{\circ}$ and near $\theta = 3.5^{\circ}$, indicating that the nature of the superconducting state may vary across this range. At $\theta \approx 5^{\circ}$, superconductivity in $t$WSe$_2$ emerges in a narrow region at finite displacement fields, closely following van Hove singularities, and is adjacent to a metallic phase exhibiting Fermi surface reconstruction driven by intervalley coherent antiferromagnetic order \cite{Guo2025Superconductivity}. In contrast, at twist angles near $3.5^{\circ}$, superconductivity appears around zero displacement field and $\nu_h=1$, away from van Hove singularities, and is adjacent to a correlated insulator at a finite displacement field \cite{Xia2025Superconductivity}. These differences suggest that superconductivity in $t$WSe$_2$ near $\theta = 5^{\circ}$ is likely driven by weak-coupling pairing mechanism arising from Fermi surface instability, whereas that near $\theta = 3.5^{\circ}$ is more likely governed by a strong-coupling mechanism. A common finding in the two experiments of Refs.~\cite{Guo2025Superconductivity} and \cite{Xia2025Superconductivity} is that superconductivity can emerge from a symmetry-unbroken normal state, suggesting pairing between the two opposite valleys that are time-reversal partners—i.e., intervalley pairing. A variety of theoretical approaches have been employed to explore the microscopic origin of superconductivity in the different twist-angle regimes \cite{Xie2025Superconductivity,Christos2025Approximate,Zhu2025Superconductivity,Kim2025Theory,guerci2024topological,qin2024kohnluttinger,fischer2024theory,tuo2024theory}. If these two limits are connected continuously in the parameter space of $\theta$, $\nu_h$, and displacement field, $t$WSe$_2$ could act as an attractive laboratory to study the crossover between superconductivity at weak and strong coupling.

The superconductivity reported in $t$MoTe$_2$ at $\theta=3.83^{\circ}$ has notably different characteristics \cite{Xu2025signatures}. It is embedded in a ferromagnetic metal and adjacent to the FQAH insulator at $\nu_h=2/3$ and the reentrant IQAH insulator.  Above the superconducting transition temperature, $t$MoTe$_2$ enters a metallic phase with anomalous Hall effect and magnetic hysteresis, indicating the emergence of a superconductivity from a valley-polarized ferromagnetic parent state. This suggests that the electron pairing is within a spin/valley polarized band. Similar superconductivity, emerging out of a ferromagnetic metal with an anomalous Hall effect and magnetic hysterisis, has also been observed in moir\'eless rhombohedral multilayer graphene, although not in proximity to a FQAH or IQAH insulator \cite{Han2025Signatures}. 

The theoretical implications of pairing within a valley-polarized, ferromagnetic state are profound. In conventional superconductors such as aluminum, time-reversal symmetry plays a central role in stabilizing spin-singlet Cooper pairing of time-reversed states. In $t$MoTe$_2$ time-reversal symmetry is already broken in the normal state, and pairing 
must occur within a single spin-polarized band, so that the 
same-spin pairing channel is the only one available.  This superconductor is 
believed to be chiral because it should have finite orbital 
magnetization for pairing within the same valley \cite{Kallin2016Chiral,Han2025Signatures}. If the superconducting state is gapped, as suggested in Ref.~\cite{xu2025chiral},
the Bogoliubov–de Gennes Hamiltonian can host bulk bands with finite Chern numbers and gapless Majorana edge modes.
The close proximity between the superconducting phase and the FQAH insulator in $t$MoTe$_2$ further raises questions about their potential microscopic connection.  In this context, the theoretical possibility of anyon superconductivity—a phase in which fractionalized quasiparticle excitations of a parent FQAH state condense into a superconducting order—has attracted growing interest \cite{shi2024doping,zhang2025holon,nosov2025anyon,shi2025anyon}.

\section{Outlook}

Twisted TMD homobilayers represent a new frontier in the study of correlated and topological phases in two-dimensional systems. The combination of strong spin-orbit coupling, moir\'e flat bands with valley contrasting topological properties, and pronounced electron-electron interactions has enabled the discovery of a remarkably rich variety of emergent quantum phases. Going forward, there are many open directions for both experiment and theory that will not only deepen our understanding of the underlying physics but may also lead to opportunities for realizing new quantum phases of matter, and for 
materials and quantum device engineering.

On the experimental front, improving device quality may uncover new
fragile and exotic quantum phases. A persistent challenge in twistronics is the 
inevitability of atomic scale disorder, twist angle inhomogeneity, and nonuniform strain—all of which can suppress energy gaps, obscure quantization, and hide symmetry-breaking phenomena. In particular, limited mobility often leads to a finite longitudinal resistance that masks the dissipationless nature expected in ideal quantum Hall states. Insights from fractional quantum Hall systems suggest that many delicate phases emerge only when mobility exceeds a critical threshold \cite{Chung2021Ultra}.  Achieving high mobility and low contact resistance in twisted TMD homobilayers is similarly crucial. Two independent transport measurements of $t$MoTe$_2$ with improved sample quality have achieved a dissipationless FQAH effect with vanishing longitudinal resistance \cite{park2025observation}, and have separately revealed superconductivity adajenct to the FQAH insulator \cite{Xu2025signatures}, underscoring the key role of device quality.

A key direction for future research is the systematic exploration and classification of the correlated and topological phases that emerge in twisted TMDs as functions of twist angle, carrier density, electric displacement field, magnetic field, and applied pressure. Thus far, only a subset of twist angles have been investigated in detail. Systematic studies may reveal new quantum states, competing orders, or phase transitions.  For example, nonabelian fractionalized states with spontaneous valley polarization have been numerically predicted in $t$MoTe$_2$ at half filling of the second moir\'e valence bands near $\theta \approx 2.0^{\circ}$ \cite{wang2025higher,xu2025multiple,Reddy2024nonabelian,Ahn2024nonabelian,Chen2025Robust}, but await experimental confirmation. Similarly, pressure has been predicted to stabilize the FQAH insulator in $t$WSe$_2$ \cite{Nicolas2023Pressure}, motivating further experimental investigation in this direction.

A major open question is the direct detection of charge fractionalization and anyonic statistics of quasiparticles in the FQAH insulator in $t$MoTe$_2$. Probing fractional charge via shot noise \cite{Saminadayar1997Observation,de-Picciotto1997Direct} or scanning tunneling measurements \cite{Papi2018Imaging}, and detecting fractional statistics through interferometry\cite{Nakamura2020Direct} or anyon collider measurements \cite{Bartolomei2020Fractional}, are essential next steps. 
Realizing such measurements in $t$MoTe$_2$ would further establish the FQAH state as a fractional topological phase and mark an important step in accessing and manipulating anyonic excitations without an external magnetic field. Moreover, if non-Abelian fractionalized states could be realized in $t$MoTe$_2$, these techniques would be indispensable for advancing toward fault-tolerant quantum computation \cite{DasSarma2005Topologically}.

Superconductivity in twisted TMD homobilayers is another fertile area for future exploration. It has been observed in both $t$WSe$_2$ and $t$MoTe$_2$, appearing in proximity to correlated states or even fractional topological phases \cite{Xia2025Superconductivity,Guo2025Superconductivity,Xu2025signatures}.
However, the nature of the superconducting state and the pairing 
mechanism remains unresolved. Experimental techniques such as tunneling spectroscopy, Josephson junction measurements, magneto-optical spectroscopy, and thermal conductivity could shed light on the pairing symmetry.  Applying in-plane magnetic fields may tune the superconducting state \cite{Xie2023Orbital,zhu2025inplane}, while angle-resolved measurements of in-plane critical currents can probe nematicity and the superconducting diode effect. Furthermore, junctions formed between superconducting regions and integer or fractional topological states—such as IQAH, IQSH or FQAH insulators—provide a promising platform for engineering exotic quasiparticles, including Majorana zero modes \cite{Luo2024}  and parafermions \cite{Lindner2012Fractionalizing}. 

Theoretical progress will be essential to guide and interpret the growing body of experimental discoveries in twisted TMD homobilayers. Continuum models have been widely used to reveal key qualitative features such as flat moir\'e bands and the emergence of topological Chern bands.  On the other hand, DFT calculations are indispensable in quantifying the role of lattice relaxation in moir\'e superlattices and its influence on the twist-angle dependence of the electronic structure. However, DFT studies of large moir\'e unit cells are computationally demanding and often require approximations or downfolding strategies to extract effective models. Nevertheless, they are crucial for validating and refining continuum models. Systematic refinement of these models—alongside ab initio-informed parameterization—remains essential to bridge theory with experiment and identify realistic conditions under which interaction-driven phenomena can emerge.

Beyond single-particle descriptions, uncovering correlated and topological phases in moir\'e TMDs requires a diverse set of many-body methods, including mean-field theories, exact diagonalization, DMRG, and quantum Monte Carlo. Each approach provides valuable, often complementary insights. A coherent understanding of IQAH and FQAH insulators is gradually emerging through these techniques. However, developing a controlled, microscopic theory of superconductivity in these systems remains a critical challenge. Addressing this challenge will require systematic benchmarking of numerical results against each other and experimental data. Additionally, careful analysis of simplified yet realistic models will be essential to isolate and identify the key physical mechanisms at play. The unique ability to tune and follow the evolution of electronic properties as twist angle and carrier density are changed could make it easier to identify trends and establish rules of thumb that apply outside of the moir\'e materials world.

In the long term, twisted TMD homobilayers could give rise to a new class of functional quantum materials, leveraging the unique capability of a single device to host a wide range of electrically tunable quantum phases.  The realization of topologically protected edge states and domain wall modes—manipulable by electric and magnetic fields—can offer a path toward gate-defined topological circuits with low dissipation. Simultaneously, the emergence of gate-tunable superconductivity enables the design of moir\'e-based Josephson junctions and hybrid platforms that combine superconducting and topological states, potentially supporting non-Abelian excitations for topological quantum computation. 
Beyond the materials themselves, the novel quantum phenomena observed in twisted TMDs can inspire the design and engineering of quantum phases across a range of platforms, including other moir\'e systems, atomic scale solid-state materials, cold atoms in optical lattices,  and quantum computing platforms such as superconducting circuits or Rydberg atoms. 

\section{Acknowledgments}
F. W. was supported by National Key Research and Development Program of China (Grants No. 2022YFA1402400 and No. 2021YFA1401300), National Natural Science Foundation of China (Grant No. 12274333). A. H. M. was supported by a Simons Targeted Grant under Award
No. 896630. 

\bibliography{ref}

\end{document}